\documentclass{article}
\usepackage{amssymb}

\input{tcilatex}
\begin{document}

\begin{center}
\textbf{Correct light deflection in Weyl conformal gravity}

\bigskip

Carlo Cattani$^{1,a}$, Massimo Scalia$^{2,b}$, Ettore Laserra$^{1,c}$, Ivana
Bochicchio$^{1,d}$

and Kamal K. Nandi$^{3,4,e}$

$\bigskip $

$^{1}$Department of Mathematics, University of Salerno, Via ponte Don
Melillo, 84084 Fisciano, Italy

$^{2}$Department of Mathematics, University of Rome "La Sapienza", P.le Aldo
Moro, Rome, Italy

$^{3}$Department of Mathematics, University of North Bengal, Siliguri 734
013, India

$^{4}$Zel'dovich International Center for Astrophysics, 3A, October
Revolution Street, Ufa 450000, Russia

$\bigskip $

$^{a}$Email: ccattani@unisa.it

$^{b}$Email: Massimo.Scalia@uniroma1.it

$^{c}$Email: elaserra@unisa.it

$^{d}$Email: ibochicchio@unisa.it

$^{e}$Email: kamalnandi1952@yahoo.co.in

\bigskip
\end{center}


\begin{center}
\textbf{Abstract}
\end{center}

The\ conformal gravity fit to observed galactic rotation curves requires $%
\gamma >0$. On the other hand, conventional method for light deflection by
galaxies gives a negative contribution to Schwarzschild value for $\gamma >0$%
, which is contrary to observation. Thus, it is very important that the
contribution to bending should in principle be \textit{positive}, no matter
how small its magnitude is. Here we show that the Rindler-Ishak method gives
a positive contribution to Schwarzschild deflection for $\gamma >0$, as
desired. We also obtain the exact local coupling term derived earlier by
Sereno. These results indicate that conformal gravity can potentially test
well against all astrophysical observations to date.

\begin{center}
--------------------------------------------------------------------------
\end{center}

The metric exterior to a static spherically symmetric distribution in Weyl
conformal gravity has been obtained by Mannheim and Kazanas [1]. Recently,
the\ solution has been used to predict rotation curves of many galaxy
samples [2] and that the model can provide a good idea of the possible size
of individual galaxies [3].The metric reads ($G=c=1$):%
\begin{eqnarray}
d\tau ^{2} &=&-B(r)dt^{2}+\frac{1}{B(r)}dr^{2}+r^{2}(d\theta ^{2}+\sin
^{2}\theta d\varphi ^{2}),\text{ \ }  \nonumber \\
B(r) &=&\alpha -\frac{2M}{r}+\gamma r-kr^{2},
\end{eqnarray}%
where $\alpha =(1-6M\gamma )^{1/2}$, $M$ is the luminous mass, $k$ and $%
\gamma $ are arbitrary constants that could be appropriately fixed by using
the fit to rotation curves. For distances neither too small nor too large,
one may take $\alpha =1$ but in what follows we shall not make any such
approximation. Now, conventional calculations for light deflection show that
the constant $k$ does not appear in the relevant equations, leading finally
to the two way deflection as [4]:%
\begin{equation}
2\epsilon =\frac{4M}{r_{0}}-\gamma r_{0},\text{ \ \ }
\end{equation}%
where $r_{0}$ is the distance of closest approach. The difficulty is that
the fit to observed rotation curve requires $\gamma >0$, and for
consistency, all other astrophysical observations should respect this sign.
Evidently, for $\gamma >0$ in Eq.(2), the light deflection by a galaxy falls
short of the Schwarzschild value $\frac{4M}{r_{0}}$, while observations tell
us that $2\epsilon >\frac{4M}{r_{0}}$. The purpose of this Brief Report is
to show that conformal gravity does give a positive contribution to
Schwarzschild deflection removing the above impass\'{e}.

The resolution is based on the realization that conventional methods do not
apply to asymptotically non-flat spacetimes as the limit $r\rightarrow
\infty $ makes no sense in it [5]. The Rindler-Ishak method of invariant
angle is most appropriate in such situations, and we show that it gives a
positive contribution to light bending proportional to $+\gamma $, as
required. The bending angle in general is defined by $\epsilon =\psi
-\varphi $. Rindler and Ishak considered the case $\varphi =0$ so that the
deflection angle is $\epsilon =\psi $ given by [5] 
\begin{equation}
\tan \psi =\frac{B^{1/2}r}{\left\vert A\right\vert },
\end{equation}%
where $A(r,\varphi )=\frac{dr}{d\varphi }$. With $u=\frac{1}{r}$, the photon
trajectory from (1) is given by%
\begin{equation}
\frac{d^{2}u}{d\varphi ^{2}}=-\alpha u+3Mu^{2}-\frac{\gamma }{2}.
\end{equation}%
As evident, $k$ has disappeared from the above equation. This is a nonlinear
differential equation that has to be solved perturbatively in powers of $M$.
Following Bodenner and Will [6], we linearize the equation by expanding $u$
in orders of $M$. To first order, we have, for small perturbation $u_{1}$: 
\begin{equation}
\frac{1}{r}=u=u_{0}+u_{1}.
\end{equation}%
Then the zeroth and first order linearized equations respectively become

\begin{equation}
\frac{d^{2}u_{0}}{d\varphi ^{2}}+\alpha u_{0}=-\frac{\gamma }{2}
\end{equation}%
\begin{equation}
\frac{d^{2}u_{1}}{d\varphi ^{2}}+\alpha u_{1}=3Mu_{0}^{2}.
\end{equation}%
In Eq.(6), redefine $\alpha u_{0}=\widetilde{u}_{0}$, $\overline{u}_{0}=%
\widetilde{u}_{0}+\frac{\gamma }{2}$, $\sqrt{\alpha }\varphi =\overline{%
\varphi }$, then it transforms into 
\begin{equation}
\frac{d^{2}\overline{u}_{0}}{d\overline{\varphi }^{2}}+\overline{u}_{0}=0,
\end{equation}%
which yields%
\begin{equation}
\overline{u}_{0}=\frac{1}{R}\cos (\overline{\varphi }).
\end{equation}%
Reverting to original variables,we get 
\begin{equation}
u_{0}=\frac{1}{\alpha }\left( -\frac{\gamma }{2}+\frac{1}{R}\cos \left\{ 
\sqrt{\alpha }\varphi \right\} \right) .
\end{equation}%
Note that a $\sqrt{\alpha }$ factor has sneaked into the argument of the
trigonometric function and also appears at other places. Their contributions
must also be included in the deflection angle. The integration of the linear
Eq.(7) can be straightforwardly performed by using standard method\footnote{%
Define the operator $D\equiv \frac{d}{d\varphi }$ and write the Particular
Integral of Eq.(7) as $u_{1}=\frac{1}{\left( D^{2}+\alpha \right) }\left[ 
\frac{C}{2}+A+B\cos \left\{ \sqrt{\alpha }\varphi \right\} +\frac{C}{2}\cos
\left\{ 2\sqrt{\alpha }\varphi \right\} \right] $ where the constants are $A=%
\frac{3M\gamma ^{2}}{4\alpha ^{2}},$ $B=-\frac{3M\gamma }{R\alpha ^{2}},$ $C=%
\frac{3M}{R^{2}\alpha ^{2}}.$ Note that $\left( D^{2}+\alpha \right) \left( 
\frac{C}{2\alpha }\right) =\frac{C}{2}\Rightarrow \frac{C/2}{\left(
D^{2}+\alpha \right) }=\frac{C}{2\alpha }=\left( \frac{M}{4R^{2}\alpha ^{3}}%
\right) \times 6$ etc. Also use $\left( D^{2}+\alpha \right) \varphi \sin
\left( \sqrt{\alpha }\varphi \right) =2\sqrt{\alpha }\cos \left( \sqrt{%
\alpha }\varphi \right) $ $\Rightarrow \frac{\cos \left( \sqrt{\alpha }%
\varphi \right) }{\left( D^{2}+\alpha \right) }=\frac{\varphi \sin \left( 
\sqrt{\alpha }\varphi \right) }{2\sqrt{\alpha }}$. Similarly, $\frac{1}{%
\left( D^{2}+\alpha \right) }\cos \left\{ 2\sqrt{\alpha }\varphi \right\} =-%
\frac{1}{3\alpha }\cos \left\{ 2\sqrt{\alpha }\varphi \right\} $. Adding the
Characteristic Function from $\left( D^{2}+\alpha \right) u_{1}=0$, we
arrive at Eq.(11).}. The solution is 
\begin{eqnarray}
u_{1} &=&\frac{M}{4R^{2}\alpha ^{3}}[6+3R^{2}\gamma ^{2}-6R\gamma \cos
\left\{ \sqrt{\alpha }\varphi \right\}  \nonumber \\
&&-2\cos \left\{ 2\sqrt{\alpha }\varphi \right\} -6R\sqrt{\alpha }\gamma
\varphi \sin \left\{ \sqrt{\alpha }\varphi \right\} ],
\end{eqnarray}%
Then the perturbative orbit equation, after changing $\varphi \rightarrow
\pi /2-\varphi $ on the right hand sides of Eqs.(10,11), is given by 
\begin{eqnarray}
u &=&u_{0}+u_{1}=\frac{1}{\alpha }\left[ -\frac{\gamma }{2}+\frac{1}{R}\cos
\left\{ \frac{\sqrt{\alpha }}{2}\left( \pi -2\varphi \right) \right\} \right]
+\frac{M}{4R^{2}\alpha ^{3}}[6+3R^{2}\gamma ^{2}  \nonumber \\
&&-6R\gamma \cos \left\{ \frac{\sqrt{\alpha }}{2}\left( \pi -2\varphi
\right) \right\} -2\cos \left\{ \sqrt{\alpha }\left( \pi -2\varphi \right)
\right\}  \nonumber \\
&&-3\pi R\sqrt{\alpha }\gamma \sin \left\{ \frac{\sqrt{\alpha }}{2}\left(
\pi -2\varphi \right) \right\} +6R\sqrt{\alpha }\gamma \varphi \sin \left\{ 
\frac{\sqrt{\alpha }}{2}\left( \pi -2\varphi \right) \right\} ].
\end{eqnarray}%
Note that the usual Schwarzschild orbit equation $u=\frac{1}{R}\sin \varphi +%
\frac{M}{2R^{2}}\left( 3+\cos 2\varphi \right) $ is recovered at $\gamma =0$
and $\alpha =1$. From Eq.(12), we can find $r$ at $\varphi =0$ as 
\begin{equation}
r=\frac{4\alpha ^{3}R^{2}}{X},
\end{equation}%
where%
\begin{eqnarray}
X &\equiv &6M-2\alpha ^{2}R^{2}\gamma +3MR^{2}\gamma ^{2}-R(6M\gamma
-4\alpha ^{2})\cos \left\{ \frac{\pi \sqrt{\alpha }}{2}\right\}  \nonumber \\
&&-2M\cos \left\{ \pi \sqrt{\alpha }\right\} -3MR\pi \sqrt{\alpha }\gamma
\sin \left\{ \frac{\pi \sqrt{\alpha }}{2}\right\} .
\end{eqnarray}%
Also, at $\varphi =0$, we find that 
\begin{equation}
\left\vert A\right\vert =\frac{4R^{2}\alpha ^{7/2}[3MR\pi \sqrt{\alpha }%
\gamma \cos \left\{ \frac{\pi \sqrt{\alpha }}{2}\right\} +4R\alpha ^{2}\sin
\left\{ \frac{\pi \sqrt{\alpha }}{2}\right\} -4M\sin \left\{ \pi \sqrt{%
\alpha }\right\} ]}{X^{2}}.
\end{equation}%
It can be seen again that, at $\gamma =0$, we recover the Schwarzschild
values $r=\frac{R^{2}}{2M}$ and $\left\vert A\right\vert =\frac{R^{3}}{4M^{2}%
}$. Using the value of $r$ from Eq.(13) and $\left\vert A\right\vert $ from
Eq.(15), we get from Eq.(3) the required deflection angle 
\begin{equation}
\tan \psi =\frac{X\sqrt{B(r)}}{3M\pi R\gamma \alpha \cos \left\{ \frac{\pi 
\sqrt{\alpha }}{2}\right\} +4\sqrt{\alpha }[R\alpha ^{2}\sin \left\{ \frac{%
\pi \sqrt{\alpha }}{2}\right\} -M\sin \left\{ \pi \sqrt{\alpha }\right\} ]}.
\end{equation}%
This is the result we get considering the exact metric without any \textit{a
priori }approximation on $u$ or $\alpha $. Restoring the value of $\alpha $,
expanding in the first power of $\gamma $ and then in first power in $R$, we
obtain for small $\psi $, after converting to $r_{0}$ via $\frac{1}{r_{0}}=%
\frac{1}{R}+\frac{M}{R^{2}}\Rightarrow R\simeq r_{0}$, the leading order
terms 
\begin{equation}
2\psi =\frac{4M}{r_{0}}-\frac{kr_{0}^{3}}{2M}+\frac{15M^{2}\gamma }{r_{0}}.
\end{equation}%
The second term is the same as the one obtained by Rindler and Ishak [4] for
the deflection in the Schwarzschild-de Sitter spacetime. Using $k=\Lambda /3$
for comparison with literature and expressing $r_{0}$ in terms of the impact
parameter $b$ as $\frac{1}{r_{0}}\simeq \frac{1}{b}+\frac{\Lambda b}{6}$, we
have the relevant terms 
\begin{equation}
2\psi =\frac{4M}{r_{0}}+\frac{15M^{2}\gamma }{r_{0}}\simeq \frac{4M}{b}+%
\frac{2Mb\Lambda }{3}+\frac{15M^{2}\gamma }{b}>\frac{4M}{r_{0}}.
\end{equation}%
Note that we have also obtained the local coupling term $\frac{2Mb\Lambda }{3%
}$ derived earlier by Sereno [5] by a completely different method, namely,
by integrating the first order differential equation of light orbit. Our
main result is that we have obtained a positive contribution $+\frac{%
15M^{2}\gamma }{b}$ instead of a negative contribution. This positivity is
important \textit{as a principle }since it lends physical consistency to
conformal gravity predictions.

Here we wish to point out that Sultana and Kazanas [7] have first tackled
the present problem of light deflection. To make contact with their
calculation, we should redefine our $M$ as 
\begin{equation}
M=\frac{\beta }{2}(2-3\beta \gamma )\Rightarrow \alpha =(1-6M\gamma
)^{1/2}=1-3\beta \gamma .
\end{equation}%
They used the path equation to first order in $\gamma $ as 
\begin{equation}
u_{\text{SK}}=\left( \frac{\sin \varphi }{b}-\frac{\gamma }{2}\right) +\left[
\frac{3\beta (2-3\beta \gamma )}{4b^{2}}+\frac{\beta (2-3\beta \gamma )}{%
4b^{2}}\cos 2\varphi \right] -\frac{3\beta \gamma }{2b}\varphi \cos \varphi
\end{equation}%
that yielded a negative contribution $-\frac{4\beta ^{2}\gamma }{b}$ to two
way deflection. However, note that in searching for the first power effect
of $\gamma $, it is only logical that one must retain all the first power
terms in $\gamma $ in relevant expansions, which in turn implies that one
must retain $\alpha \neq 1$ in the trigonometric arguments and elsewhere.
Then the expression for $u$ from Eq.(12), to first order in $\gamma $, reads 
\begin{eqnarray}
u &=&\left( \frac{\sin \varphi }{b}-\frac{\gamma }{2}\right) +\frac{3\beta
\gamma }{2b}\sin \varphi  \nonumber \\
&&+\left[ \frac{3\beta (2+15\beta \gamma )}{4b^{2}}+\frac{\beta (2+15\beta
\gamma )}{4b^{2}}\cos 2\varphi +\frac{3\beta ^{2}\gamma }{4b^{2}}(2\varphi
-\pi )\sin 2\varphi \right] ,
\end{eqnarray}%
which is widely different from $u_{\text{SK}}$. Thus the negative
contribution seems ruled out. To see the actual contribution, it is enough
to convert $2\psi =\frac{4M}{r_{0}}+\frac{15M^{2}\gamma }{r_{0}}$ in terms
of the notation $\beta $ used in Ref.[7], which would then yield, to first
order in $\gamma $, the result $2\psi =\frac{4\beta }{r_{0}}+\frac{9\beta
^{2}\gamma }{r_{0}}$. Yet again, the positive $\gamma -$contribution is
quite evident.

We can incorporate the light bending Eq.(18) in the lensing equation
ignoring the local coupling term, which is numerically much smaller than the 
$\gamma $ term by several orders of magnitude for typical galaxies. The lens
equation is given by%
\begin{equation}
\theta D_{os}=\beta D_{os}+(2\psi )D_{ls}.
\end{equation}%
When the observer, lens and source are aligned in one direction, we have $%
\beta =0$, which yields, in the small angle approximation $b=\theta D_{ol}$,
the "Weyl angle" as 
\begin{equation}
\theta _{\text{Weyl}}=\left( \frac{4M+15M^{2}\gamma }{D}\right) ^{1/2},
\end{equation}%
where $D\equiv \frac{D_{ol}D_{os}}{D_{ls}}$. The Einstein angle is of course 
$\theta _{\text{Einstein}}=\left( \frac{4M}{D}\right) ^{1/2}$, which means
that the Schwarzschild mass is only to be redefined as $\overline{M}=M+\frac{%
15}{4}M^{2}\gamma $ to obtain the Weyl angle. Of course, for galactic
lenses, these masses do not differ enormously. This is expected as the
luminous matter obeys $M_{\text{L}}(r)\varpropto r^{3}$, while flat rotation
curves demand $M_{\text{DM}}(r)\varpropto r$ \ in the halo region that
increases with radius more slowly with distance than $M_{\text{L}}(r)$ and
thus is comparatively rarer\footnote{%
We are thankful to Prof. Maria Assunta Pozio for pointing this out.}.
Customarily, the observed light deflection is explained by a total mass
distribution that includes also the hypothetical "dark matter" in \textit{%
excess }of the luminous component [4]. On the other hand, in conformal
gravity, this hypothesis is not required [1-3]. Our result here supports
this central aspect of conformal gravity in that the mass gets \textit{%
automatically} enhanced to $\overline{M}>M$ due necessarily to the positive
contribution $+\frac{15}{4}M^{2}\gamma $, as we promised to show.

It has been pointed out to us that the $\gamma r$ term can be absorbed into
a conformal factor [8] and that the sign of the $\gamma -$contribution to
bending can alter under different choices of conformal factors [9]\footnote{%
We thank an anonymous referee for pointing these out.}. While we agree with
these facts, we still worked only in the conformal frame as exactly fixed by
Eq.(1), i.e., in the metric used by Sultana and Kazanas [7], for the single
reason that it has remarkably explained observations for appropriate choices
of $\gamma $ and a quadratic potential [2]. It is true that the bending
effect is exceedingly small and, as it stands, incompatible with the
observations but our aim was to argue that the sign must be positive in the
first place for qualitative validity of conformal gravity theory.\footnote{%
Under the practical assumption that $R>>M$, $\frac{1}{R}\simeq \frac{1}{r_{0}%
}\simeq \frac{1}{b}+\frac{\Lambda b}{6}$, the $\gamma -$bending is always
positive though exceedingly small.} Thus, it remains on us to explore if the
theory can show also quantitative validity in respect of bending
observations. Work is underway.

\textbf{Acknowledgment}

The authors thank Manoranjan Singha for helpful discussion. KKN acknowledges
financial support from the Erasmus Grant administered by the University of
Salerno, Italy.

\textbf{References}

[1] P. D. Mannheim and D. Kazanas, Astrophys. J. \textbf{342}, 635 (1989).

[2] P.D. Mannheim and J.G. O'Brien, Phys. Rev. Lett. \textbf{106}, 121101
(2011).

[3] K.K. Nandi and A. Bhadra, Phys. Rev. Lett. \textbf{109}, 079001 (2012).

[4] A. Edery and M. B. Paranjape, Phys. Rev. D \textbf{58}, 024011 (1998).

[5] W. Rindler and M. Ishak, Phys. Rev. D \textbf{76}, 043006 (2007).

[6] J. Bodenner and C.M. Will, Am. J. Phys. \textbf{71}, 770 (2003).

[7] J. Sultana and D. Kazanas, Phys. Rev. D \textbf{81}, 127502 (2010).

[8] P. D. Mannheim and D. Kazanas, Phys. Rev. D \textbf{44}, 417 (1991).

[9] A. Edery \textit{et al}, Gen. Rel. Grav. \textbf{33}, 2075 (2001).

\end{document}